\documentclass[prb,aps,preprint,superscriptaddress,groupaddress]{revtex4}
\usepackage{amssymb}
\usepackage{latexsym}
\usepackage{amsmath}
\usepackage[mathcal]{eucal}
\usepackage{graphicx}
\usepackage{dcolumn}
\usepackage{amsfonts}
\usepackage{bm}
\usepackage{epsfig}
\usepackage{array}

\newcommand{\be}{\begin{equation}}
\newcommand{\ee}{\end{equation}}
\newcommand{\bea}{\begin{eqnarray}}
\newcommand{\eea}{\end{eqnarray}}

\renewcommand{\vec}[1]{{\bm #1}}

\bibpunct{[}{]}{,\!}{n}{,}{,}

\begin{document}
\title{Inter-pocket pairing and gap symmetry in Fe-based superconductors with only electron pockets}
\author{M. Khodas}
\affiliation{Department of Physics and Astronomy, University of Iowa, Iowa City, Iowa 52242, USA}
\author{A. V. Chubukov}
\affiliation{Department of Physics, University of Wisconsin, Madison, Wisconsin 53706, USA}

\date{\today }

\begin{abstract}
Pairing symmetry in recently discovered Fe-based metallic superconductors  AFe$_2$Se$_2$ (A = K, Rb, Cs)
 with high transition temperature
 $T_c \sim 40$ K
is currently a subject of intensive debates.
 These systems contain only electron pockets, according to photoemission, and differ from the majority of Fe-based
 superconductors in which both electron and hole pockets are present.
 Both $d$-wave and $s$-wave pairing symmetries
  have been proposed for AFe$_2$Se$_2$, but a $d$-wave gap
 generally has nodes, while experiments clearly point to no-nodal behavior, and a conventional
  $s$-wave gap  is inconsistent with the observation of the neutron resonance below $T_c$.
   We argue that current theories of  pairing in such systems are incomplete and must include
  not only  intra-pocket pairing condensate but also
   inter-pocket condensate made of fermions belonging to different electron pockets. We
  analyze the interplay between intra-pocket and inter-pocket pairing depending on the ellipticity of electron pockets and the strength of their hybridization and show that
  hybridization brings the system 
     into a new $s^{+-}$ state, in which the gap changes sign between hybridized pockets. This state has the full gap and at the same time
      supports spin resonance,  in agreement with the data.
     Near the boundary of $s^{+-}$ state we found a long-thought $s+id$ state which breaks time-reversal symmetry. 
 \end{abstract}

\maketitle

{\it {\bf Introduction:}}~~~High-temperature superconductivity in Fe-pnictides has been discovered in 2008 (Ref. ~\cite{discovery}) and almost instantly made to the top of the list
 of the most relevant issues for the physics community~\cite{review,review_1,review_2,review_3}.
   The interest in Fe-pnictides and  Fe-chalcogenides, which display similar behavior, is two-fold. On one hand, these materials hold a
  strong potential for practical applications, on the other they display superconducting properties  which does not fit into conventional
   classification scheme of superconductivity. In this scheme, $s$-wave superconductors have a gap of one sign and no nodes, $d$-wave superconductors have
    a sing-changing gap and line nodes (unless a $d$-wave state breaks time-reversal symmetry~\cite{rahul}), etc.
    For Fe-based superconductors (FeSCs),  photoemission, penetration depth, thermal conductivity, neutron scattering,
     tunneling, and other experiments done on weakly/moderately doped systems show that  the gap is $s$-wave, yet it very likely  changes sign, and even has nodes in some materials.

 The  sign-changing $s$-wave superconductivity
    is generally understood as the consequence of
  complex geometry of the Fermi surface (FS),
   which in FeSCs consists of hole and electron pockets located in
  separate regions of the Brillouin zone, and of the
  competition between intra-pocket and  inter-pocket
  interactions, both originating
  from electron - electron Coulomb repulsion.
   Intra-pocket
       repulsion  is detrimental to an $s$-wave superconductivity,
    but inter-pocket repulsion
     supports $s$-wave  superconductivity in which the gap changes sign between hole and electron pockets.
    When inter-pocket interaction wins, the system develops  a  "plus-minus" $s$-wave gap~\cite{mazin_1,kuroki_1}
   when intra-pocket interaction wins, $s$-wave superconductivity still develops, but the gap acquires
   variations along electron pockets to minimize the effect of intra-pocket repulsion~\cite{review,cvv}, and develops nodes when
      these variations become large.

    This scenario works reasonably well for weakly/moderately doped FeSCs, in which both hole and electron pockets are present. Recently, however,
    superconductivity with rather high
    $T_c \sim 40$~K
    has been discovered~\cite{exp:AFESE,latest} in
 A$_{x}$Fe$_{2-y}$Se$_2$ (AFe$_2$Se$_2$) with  A = K, Rb, Cs,  which have only electron pockets, according to photoemission~\cite{exp:AFESE_ARPES}.
     The electronic states near would be hole pockets have energy of at least $60$~meV and are unlikely to contribute to superconductivity, i.e., the reasoning
      based on the interaction between hole and electron pockets cannot be applied to these systems.

Superconductivity in AFe$_2$Se$_2$  has been studied recently by several groups~\cite{graser_11,das,dhl_AFESE,maiti,mazin,si_11,bernevig_11,kontani_last}. In the Fe-only Brillouin zone, electron pockets are located near $(0,\pi)$ and $(\pi,0)$ (here and below we set interatomic spacing $a=1$).  If we just replace intra-pocket hole-electron interaction by  intra-pocket interaction between electron pockets, we find that a gap  must change sign  between $(0,\pi)$ and $(\pi,0)$ pockets. Such a gap is  antisymmetric with respect to interchange between $X$ and $Y$ coordinates
 and hence has $d$-wave symmetry~\cite{graser_11,das,dhl_AFESE,maiti}. Taken at a face value, this gap is of "plus-minus" type and has no nodes. However, no-nodal $d$-wave gap is rather fragile and was argued~\cite{mazin} to
   acquire symmetry-related nodes once one includes the hybridization between the electron pockets due to an additional interaction via a Se. 
  The data on AFe$_2$Se$_2$,
   however, show
   that the gap has no nodes~\cite{Zhang,Zhou,Wang}.
  Several groups argued~\cite{mazin,si_11,bernevig_11,kontani_last} that
  a no-nodal behavior
   indicates
  that the gap in AFe$_2$Se$_2$
   must be $s$-wave, but a conventional $s$-wave gap is
   inconsistent with recent observation of the spin resonance below $T_c$ in Rb$_x$Fe$_{2-y}$Se$_2$
(Ref.~\cite{resonance}).

In this communication, we show that nodeless superconductivity
 consistent with the spin resonance
  in fact appears quite naturally in a situation when only electron pockets are present.
   We argue that complete theory  of superconductivity in such geometry should include on equal footings a 
 pairing condensate made out of fermions on the same pocket (intra-pocket pairing) and a pairing condensate made out of fermions on different pockets (inter-pocket pairing).
  Inter-pocket pairing has been earlier discussed  for other systems, most recently
 for inter-valley pairing in graphene~\cite{efetov}. It was discussed in early days of Fe-pnictides
  regarding
 a possible spin-triplet, even-parity pairing in weakly doped FeSCs~\cite{Zhang_1,mazin_3},  but
  was not considered in previous works on the pairing in FeSCs with only electron pockets.
   We argue that,
   whenever
   two FS pockets cross upon folding and split due to hybridization,
    inter-pocket pairing must be included on equal footings with intra-pocket pairing.  
   For AFe$_2$Se$_2$  inter-pocket pairing is particularly important because both hybridization {\it and} ellipticity
    are small~\cite{mazin}, but we argue that it is generally relevant to the pairing in the presence of hybridization because intra-pocket pairing
     for non-hybridized fermions becomes inter-pocket pairing for hybridized fermions, and vise versa (see below).

   We show that the interplay between intra- and inter-pocket pairing in the standard model leads to a competition
   between $d$-wave and $s$-wave states.  We find three phases, depending on the degree of eccentricity of electron pockets and the strength of the hybridization -- an $s$-wave, a
    $d\pm i s$ state which breaks time-reversal symmetry, and a
   $d$-wave state. In $s$-wave and $s+id$ states, all states are gapped. In a $d$-wave state, there are nodes, but unusual ones -- they
    form vertical loops centered  $k_z = \pi/2$. In some range of parameters, loops collapse and a $d$-wave state also becomes nodeless.
   The $s$-wave is of plus-minus type -- the gaps on hybridized FSs have opposite signs. Such state has been earlier
   proposed phenomenologically by  I. Mazin~\cite{mazin}. Our results explain microscopic mechanism of such $s^+-$ superconductivity.

\begin{figure}[h]
\begin{center}
\includegraphics[width=0.9\columnwidth]{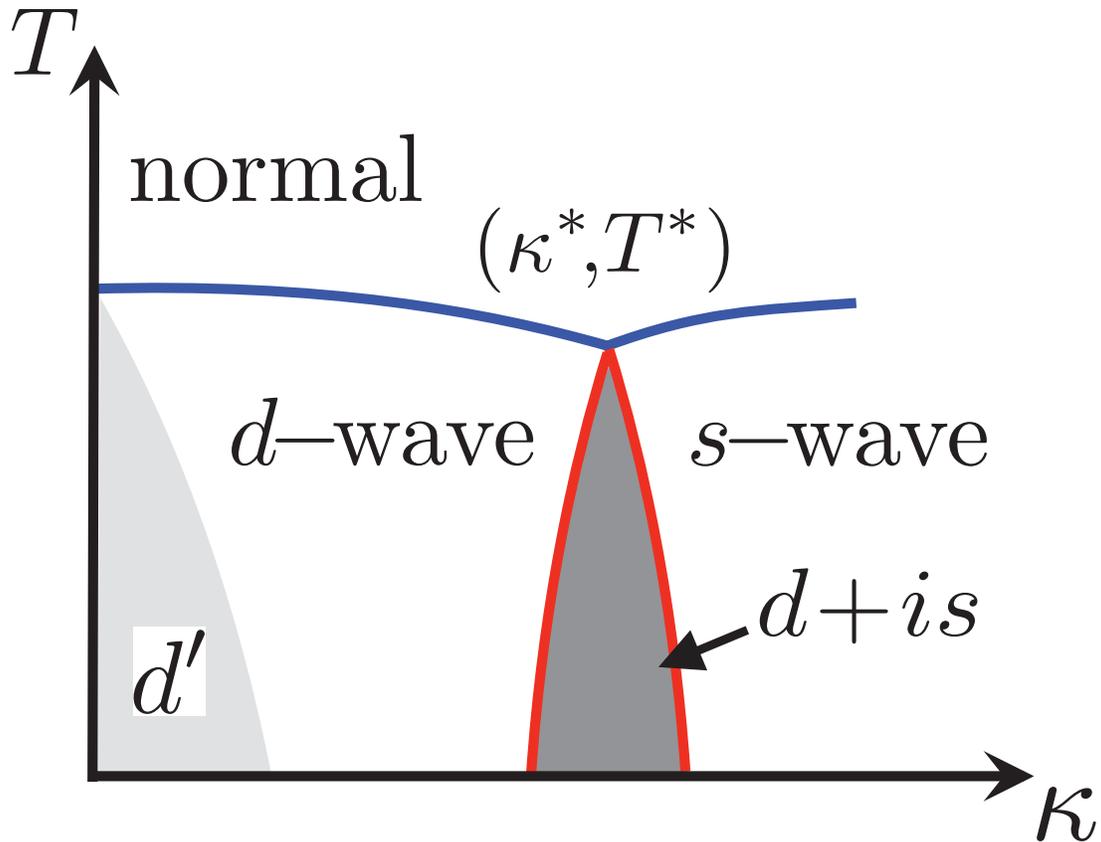}
\caption{The phase diagram in  $(\kappa,T)$--plane for  Fe-based superconductors with only electron pockets.
Two solid and almost horizontal lines (blue) separates normal and superconducting phases.
 Four solid lines of second order phase transition converge at the tetra-critical point $(\kappa^*,T^*)$.
The  $s+id$ phase with broken time reversal symmetry is shown by the dark (grey) shaded area confined by two solid (red) almost vertical lines.
The two neighboring superconducting phases at $\kappa < (>) \kappa^*$ have $d(s)$-wave ordering respectively.
In the $d'$  region  the excitation spectrum is fully gapped
 even though the order parameter has a $d$-wave symmetry.
}
\label{fig:1}
\end{center}
\end{figure}

{\it {\bf Methods:}}~~~We consider the low-energy physics of FeSCs with only electron pockets within a 2D model of
fermions near $(0,\pi)$ and $(\pi,0)$, interacting either directly,
 or via a pnictogen/chalcogen.
 The direct interaction is a momentum-conserving process, up to a lattice momentum
  in the Fe-zone, $2\pi$,  the indirect one has excess momentum
  $\vec{Q} =(\pi, \pi)$ taken by pnictide/chalcogenide.
   This last interaction then hybridizes the two electron pockets and also
   gives rise to additional 4-fermion interactions with excess momentum
   $\vec{Q}$. We show that the hybridization is overly relevant while additional interaction does not play much role.  The  hybridization in AFe$_2$Se$_2$
   actually involves momentum $(\pi,\pi,\pi)$ because of body-centered tetragonal structure of these materials, i.e.,
   hybridized fermions belong to different planes separated by $k_z =\pi$~\cite{mazin,inosov}.
To simplify
the presentation, we first consider hybridization for  a simple tetragonal
structure, for which hybridized fermions have the same $k_z$, and then extend the analysis to body-centered tetragonal structure.

Let $c^{\dag}_{\vec{k}}$ be a creation operator for electrons at $(0,\pi)$, and $f^{\dag}_{\vec{k}} = c^{\dag}_{\vec{k}+\vec{Q}}$ is a creation operator of electrons at
$(\pi,0)$.
The quadratic part of the Hamiltonian $H=H_2 +H_{int}$ is
\begin{align}\label{11}
H_2 = \sum_{\vec{k}}  \epsilon^c_{\vec{k}} c_{\vec{k}}^{\dag} c_{\vec{k}}
+
\sum_{\vec{k}} \epsilon^f_{\vec{k}} f_{\vec{k}}^{\dag} f_{\vec{k}}
+
\sum_{\vec{k}} \lambda \left[ c_{\vec{k}}^{\dag} f_{\vec{k}} + f^{\dag}_{\vec{k}} c_{\vec{k}} \right]\, ,
\end{align}
 where the first two terms describe fermion dispersion, and the last term describes the hybridization, i.e.,
the process with excess momentum $\vec{Q}$ taken by Se.
 The two elliptical FSs are defined by $\epsilon^{c(f)}_{\vec{k}} = \epsilon_F$.
We approximate
 fermion
excitations near these FSs by
\be
\epsilon^{c}_{\vec{k}} = v_F (\phi) (k-k_F( \phi)), ~~ \epsilon^{f}_{\vec{k}} = v_F (\phi+\pi/2) (k-k_F( \phi+\pi/2))\, ,
\label{12}
\ee
 where $\phi$ is the angle along each of the the FSs counted from the $x-$axis.  By virtue of tetragonal symmetry, $v_F (\phi) = v_F (1 +a \cos 2 \phi)$ and $k_F (\phi) = k_F (1 + b \cos 2 \phi)$. The anisotropy
  of the Fermi velocity does not play a major role in our analysis, but the eccentricity of the FS (the parameter $b$) is overly relevant. Both $b$ and $\lambda/(v_F k_F)$ are
  small for
  AFe$_2$Se$_2$ (Ref.\cite{mazin}), but their ratio $\kappa =
  \lambda / (v_F k_F |b|)$ can be arbitrary.
 For simplicity, we assume that $\lambda$ is independent on the direction of $\vec{k}$.  A more complex form of $\lambda (\vec{k})$ will affect our results quantitatively but not qualitatively.

The interaction Hamiltonian generally involves direct, momentum-conserving,  4-fermion interactions, and interactions via a pnictide or a chalkogenide (Se in our case) with excess momentum $\vec{Q}$.
The four direct interactions are
\begin{align}\label{14}
H_1 & = \frac{u_1}{2}
\int d \vec{x}
\left(  c_{\sigma}^{\dag} f_{\sigma'}^{\dag} f_{\sigma'} c_{\sigma}
+ f_{\sigma}^{\dag} c_{\sigma'}^{\dag} c_{\sigma'} f_{\sigma}
\right)
\notag \\
H_2 & = \frac{u_2}{2}
\int d \vec{x}
\left(
  c_{\sigma}^{\dag} f_{\sigma'}^{\dag} c_{\sigma'} f_{\sigma}  +
   f_{\sigma}^{\dag} c_{\sigma'}^{\dag} f_{\sigma'} c_{\sigma}
   \right)
\notag \\
H_3 & =
\frac{u_3}{ 2 }
\int d \vec{x}
\left( c_{\sigma}^{\dag} c_{\sigma'}^{\dag} f_{\sigma'}f_{\sigma}   + f_{\sigma}^{\dag} f_{\sigma'}^{\dag} c_{\sigma'} c_{\sigma} \right)
\notag \\
H_4 & =
\frac{u_4}{2}
\int d \vec{x}
\left( c_{\sigma}^{\dag} c_{\sigma'}^{\dag} c_{\sigma'} c_{\sigma}
+  f_{\sigma}^{\dag} f_{\sigma'}^{\dag} f_{\sigma'} f_{\sigma}  \right)
\end{align}
$H_1$ and $H_2$ are inter-band density-density and exchange interactions,
interaction, $H_4$ is the intra-band density-density interaction, and
 $H_3$ describes the umklapp pair-hopping processes.  We assume that all interactions are repulsive and angle-independent.  The interaction with excess momentum $\vec{Q}$ is
\begin{equation}
H_{\vec{Q}} = w_1 \int d \vec{x} ( c^{\dag}_{\sigma} f_{\sigma} + f^{\dag}_{\sigma}  c_{\sigma}  )( c^{\dag}_{\sigma'}  c_{\sigma'}  + f^{\dag}_{\sigma'}  f_{\sigma'} ) \, .
\end{equation}
Other interactions with $\vec{Q}$ vanish without time-reversal symmetry breaking.

The quadratic Hamiltonian can be diagonalized by unitary transformation to new operators
$a_{\vec{k}} = c_{\vec{k}} \cos \theta_{\vec{k}} + f_{\vec{k}}\sin \theta_{\vec{k}} $,
$b_{\vec{k}} = - c_{\vec{k}} \sin \theta_{\vec{k}}  + f_{\vec{k}} \cos \theta_{\vec{k}} $ with
\begin{equation}
\label{13}
\sin 2 \theta_{\vec{k}} = \frac{\lambda}{ \sqrt{ \lambda^2 + ( \epsilon^c_{\vec{k}} - \epsilon^f_{\vec{k}})^2/4 }}\,~,~\,\,\,~\cos 2 \theta_{\vec{k}} =  \frac{( \epsilon^c_{\vec{k}}- \epsilon^f_{\vec{k}})}{2 \sqrt{ \lambda^2 + ( \epsilon^c_{\vec{k}} - \epsilon^f_{\vec{k}})^2/4 }} \, .
\end{equation}
In terms of new operators,
\begin{equation}
H_2 = \sum_{\vec{k}} E_{\vec{k}}^a a_{\vec{k}}^{\dag} a_{\vec{k}}
+  \sum_{\vec{k}} E_{\vec{k}}^b b_{\vec{k}}^{\dag} b_{\vec{k}}
\end{equation}
with
\begin{equation}
E_{{\vec{k}}}^{a,b}  = \frac{ 1 }{ 2 } \left( \epsilon_{\vec{k}}^c + \epsilon_{\vec{k}}^f \right) \pm
 \left[ \lambda^2 +  \left( \epsilon_{\vec{k}}^c - \epsilon_{\vec{k}}^f \right)^2/4 \right]^{1/2}
 \, .
\end{equation}
 In our notations, $(\epsilon_{\vec{k}}^c + \epsilon_{\vec{k}}^f)/2 \approx \epsilon_F + v_F (k-k_F) = \epsilon_F + \xi$, and  $(\epsilon_{\vec{k}}^c - \epsilon_{\vec{k}}^f)/2 \approx
v_F k_F b \cos 2 \phi$, such that
$ E_{{\vec{k}}}^{a,b} -\epsilon_F  = \xi \pm \lambda \left(1 + \cos^2 2\phi/\kappa^2\right)^{1/2}$,
$\cos^2 2 \theta = \cos^2 2\phi/(\kappa^2 + \cos^2 2\phi)$, and  $\sin^2 2 \theta = \kappa^2/(\kappa^2 + \cos^2 2\phi)$.

\begin{figure}[h]
\begin{center}
\includegraphics[width=0.9\columnwidth]{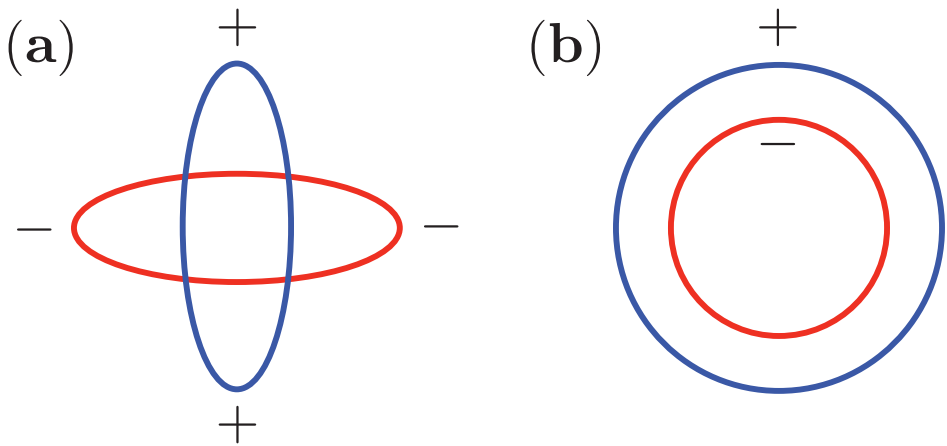}
\caption{The structure of superconducting gap at small and large $\kappa$, which is the ratio of the hybridization and the degree of ellipticity of the electron pockets. At the smallest $\kappa$ (panel a), the gap has different sign on the original FS pockets and is $d$-wave because it is antisymmetric with respect to rotation by
$90^{\circ}$.
At large $\kappa$ (panel b), the gap again changes sign, but now between hybridized FS pockets. This gap is symmetric with respect to $90^{\circ}$ rotation and is therefore an $s$-wave.}
\label{fig:2}
\end{center}
\end{figure}

{\it Qualitative reasoning:} ~~~The interplay between intra-pocket and inter-pocket pairing, and between $d$-wave and $s$-wave gap symmetry can be understood by considering the limits of small and large $\kappa$
 (Fig. \ref{fig:2}).  At $\kappa \to 0$ the hybridization vanishes, i.e., $c$ and $f$ are primary operators.
For elliptical pockets,  intra-pocket Cooper susceptibility is the largest one, and when  $u_3 > u_4$, the system
  develops a conventional pairing instability with $\Delta_c = \langle c_{\uparrow}^{\dag} c_{\downarrow}^{\dag} \rangle$ and
$\Delta_f =  \langle f_{\uparrow}^{\dag} f_{\downarrow}^{\dag} \rangle$ and $\Delta_f  =- \Delta_c$
  (Fig. \ref{fig:2}a). This solution is antisymmetric with respect to $c \leftrightarrow f$  and hence is $d$-wave. Consider next the opposite limit of large $\kappa$. Now $a$ and $b$ are primary fermion operators, and the FSs of $a$ and $b$ fermions are well separated in the momentum space
  (Fig. \ref{fig:2}b). The leading pairing instability is again a conventional intra-pocket one,  and the gaps $\Delta_a = \langle a_{\uparrow}^{\dag} a_{\downarrow}^{\dag} \rangle$ and $\Delta_b = \langle b_{\uparrow}^{\dag} b_{\downarrow}^{\dag} \rangle$ obey $\Delta_a = -\Delta_b$.
 This gap, however, is a sign-changing
 $s$-wave rather than $d$-wave.  To see this, we note that at large $\kappa$, $a_{\uparrow}^{\dag} a_{\downarrow}^{\dag} - b_{\uparrow}^{\dag} b_{\downarrow}^{\dag} = c_{\uparrow}^{\dag} f_{\downarrow}^{\dag} + f_{\uparrow}^{\dag} c_{\downarrow}^{\dag}$, i.e.,  the solution $\Delta_a = - \Delta_b$ corresponds to non-zero $ \langle c_{\uparrow}^{\dag} f_{\downarrow}^{\dag} + f_{\uparrow}^{\dag} c_{\downarrow}^{\dag} \rangle$.
 The latter combination is symmetric with respect to
 $c \leftrightarrow f$
 and hence  is an $s$-wave.
 We also see that,
  in terms of original fermions, the pairing condensate is now the
 inter-pocket
one -- it is made out of fermions belonging to different pockets.
   What happened with the $d$-wave solution?  At large $\kappa$ we have  $c_{\uparrow}^{\dag} c_{\downarrow}^{\dag} - f_{\uparrow}^{\dag} f_{\downarrow}^{\dag} = -(a_{\uparrow}^{\dag} b_{\downarrow}^{\dag} + b_{\uparrow}^{\dag} a_{\downarrow}^{\dag})$. Hence,  in terms of $a$ and $b$  operators,  $d$-wave pairing now becomes inter-pocket pairing.
 We see therefore that intra-pocket pairing in terms of one set of fermions corresponds to inter-pocket pairing in terms of the other set.
 Since the pairing symmetry changes from a $d$-wave to an $s$-wave between
   small and large $\kappa$, and
    because
   within the
 original
set
the former is
an intra-pocket  pairing and
the latter is an inter-pocket pairing,
     one {\it must} include  the two pairings on on equal footing in order  to describe the transformation from $d$- to $s$-wave
 symmetry.

 It is natural to analyze the pairing in terms of new $a$ and $b$ fermions
 because the Hamiltonian, Eq.~\eqref{11},
 is then quadratic at all values of  $\kappa$.  We introduce
  intra-- and inter--band pair creation operators,
\begin{align}\label{15}
J^{\dag}_{\pm} =\frac{ 1 }{2 } \left( a^{\dag} a^{\dag} \pm b^{\dag}b^{\dag} \right)\, , \,\,\,\,
\tilde{J}^{\dag}_{\pm} =\frac{ 1 }{2 } \left( a^{\dag} b^{\dag} \pm b^{\dag}a^{\dag} \right) \, .
\end{align}
The combinations
$J^\dagger_+$ and ${\tilde J}^\dagger_-$ describe an ordinary, "plus-plus" $s$-wave pairing and
 spin-triplet,
  even parity inter-band pairing, respectively
  (the triplet channel is identical to the one considered in~\cite{Zhang_1}).
   In our case, these two pairing channels are
   strongly repulsive, and we can safely omit them.
 The linear combinations of
  the other two components
  $J^\dagger_-$ and ${\tilde J}^\dagger_+$
 describe $s$-wave
 pair creation operators
\begin{equation}\label{16}
\frac{1}{ 2} \left( c^{\dag}_{\sigma} f^{\dag}_{\sigma'} + f^{\dag}_{\sigma} c^{\dag}_{\sigma'} \right)
 =
\left[ \cos 2 \theta \tilde{J}_+^{\dag} + \sin 2 \theta J_-^{\dag}\right]_{\sigma \sigma' }
\end{equation}
and $d$-wave pair creation operators
\begin{equation}\label{17}
\frac{1}{ 2} \left(  c^{\dag}_{\sigma} c^{\dag}_{\sigma'}  - f^{\dag}_{\sigma} f^{\dag}_{\sigma'} \right)
=
\left[ \cos 2 \theta J_-^{\dag} - \sin 2 \theta \tilde{J}_+^{\dag}\right]_{\sigma \sigma' } \, ,
\end{equation}
where the  angle $\theta$ is defined in Eq.~\eqref{13}.
 The interaction, Eq.~\eqref{14},  can then be decomposed into an
$s$-wave and $d$-wave channels, $H_{int} = H_s + H_d$ with
\begin{align}\label{Hs_eff}
H_s & =  -2 u_{s} [ s' J_-^{\dag}  + c' \tilde{J}_+^{\dag} ]_{\sigma \sigma'} [ s J_-  + c \tilde{J}_+ ]_{\sigma' \sigma} \, ,
\end{align}
\begin{align}\label{Hd_eff}
H_d & =  -2 u_{d} [ c' J_-^{\dag}  - s' \tilde{J}_+^{\dag} ]_{\sigma \sigma'} [ c J_-  - s \tilde{J}_+ ]_{\sigma' \sigma}   \, ,
\end{align}
where $c \equiv \cos 2 \theta$, $c' \equiv \cos 2 \theta'$, $s \equiv \sin 2 \theta$, $s' \equiv \sin 2 \theta'$
 and
 $ 2 u_{s} = -u_{1} - u_{2}$, $ 2 u_{d} = u_{3} - u_{4}$.
 We emphasize  that the intra-- and inter-- band pairings enter Eqs.~\eqref{Hs_eff}, \eqref{Hd_eff} on equal footing.
 We also emphasize that $J^\dagger_-$ describes a "plus-minus" s-wave pairing (different sign of the gaps on $a$ and $b$ FSs), hence our s-wave state is sign-changing s-wave.
   For circular electron pockets, the $O(2)$ rotational
  symmetry in
 $(c,f)$ space along with $SU(2)$ spin symmetry
 requires~\cite{comm}
   $u_s = u_d =u$. For weak
   ellipticity,
   $u_d$ and $u_s$
   do not have to be identical, but remain close and we keep
   $u_s = u_d =u$ for simplicity.
   A  positive $u$ is required for superconductivity.
     The interaction $H_Q$ couples these two channels with plus--plus $s$-wave channel and spin-triplet channels
     which we already neglected,
     and does not play a role in our analysis.

{\it Ginzburg-Landau Functional:}~~
To map a phase diagram in $(\kappa,T)$-plane we derive the Ginzburg-Landau Functional (GLF).
 We introduce order parameters $\Delta_s$ and $\Delta_d$ to decouple the interaction in two Cooper channels
 using the  Hubbard-Stratonovitch identity,
integrate over fermion fields and expand the effective action in powers of $\Delta_s$ and $\Delta_d$.
 Carrying out the calculations (see~\cite{suppl})
 we obtain
\begin{align}\label{GL1}
F_{GL}
=  & A_s |\Delta_s|^2 + A_d |\Delta_d|^2
+  \frac{B_s}{2}    |\Delta_s|^4 +  \frac{B_d}{2}   |\Delta_d|^4
\notag \\
&
+C |\Delta_s|^2 |\Delta_d|^2
+  \frac{E}{2} \left((\Delta_s \Delta^*_d )^2 +(\Delta^*_s \Delta_d )^2 \right) \, .
\end{align}

\begin{figure}[h]
\begin{center}
\includegraphics[width=0.96\columnwidth]{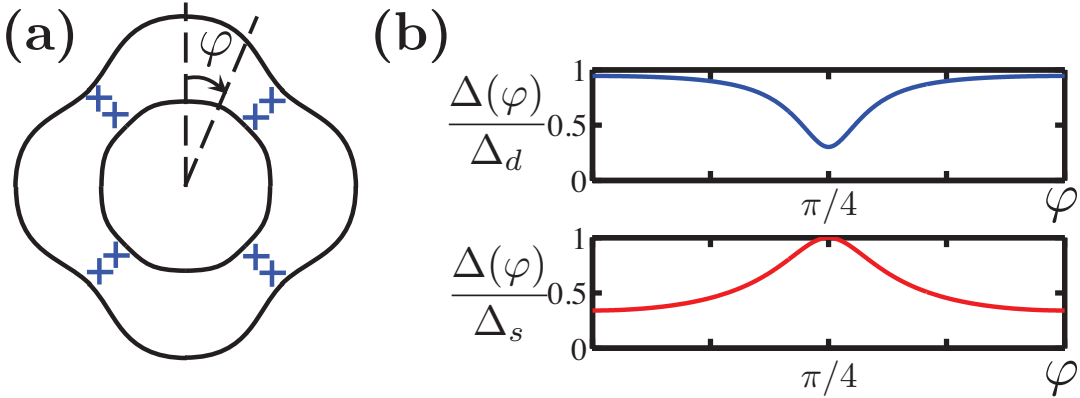}
\caption{a) The hybridized FSs at intermediate $\kappa = O(1)$. Crosses mark the location of the nodal points (zero energy states) in the $d$-wave phase. Upon approaching the boundary of $d'$ phase in Fig.\ref{fig:1}, nodal points come closer and eventually collapse, leading to a $d$-wave state with a full gap. b) The modulation of the gap magnitude along the FSs for $d$-wave and $s$-wave states.
Within our approximations, gap modulations along the two FSs are equivalent.}
\label{fig:3}
\end{center}
\end{figure}

The transition to either $s$-wave or $d$-wave state is determined by $A_s=0$ or $A_d =0$, whichever comes first.
   The lines $A_s=0$ and $A_d=0$ cross at some critical $\kappa^*$, at which $T_c = T^*_c$.
   The value of $\kappa^*$  depends on $T^*_c/\lambda$. In AFe$_2$Se$_2$, $\lambda$ is a fraction of $\epsilon_F$ and $T^*_c \ll \epsilon_F$, hence $T^*_c \ll \lambda$. In this limit, we obtained   $\kappa^* = 1/\sqrt{3}$ (in the other limit, which we consider for completeness in ~\cite{suppl}, $\kappa^* = 1/\sqrt{2}$).
   The transition temperature is determined by $\log \Lambda/T^*_c = 2/(u N_F)$,
   where $N_F = k_F/(2\pi v_F)$
    is the density of states at the
  Fermi level and $\Lambda \sim \epsilon_F$ is the upper cutoff. Near the critical $\kappa$, the first instability occurs at $T_{c,s} = T^*_c (1 + \alpha (\kappa - \kappa^*))$ for $\kappa > \kappa^*$ and at $T_{c,d} = T^*_c (1 + \alpha (\kappa^* - \kappa))$ for $\kappa < \kappa^*$, where $\alpha = 3\sqrt{3} / (2 u N_F)$ (see Fig.~\ref{fig:1}).

The
 transition from
  a $d$-wave order at $\kappa < \kappa^*$
  to an $s$-wave order at $\kappa > \kappa^*$
 is determined by the
     the interplay between fourth-order terms in  Eq. \eqref{GL1}.
      The transition can be either first order or
      continuous, via an intermediate phase where both orders are present. At $T= T^*_c$ we obtained
 $B_s = B_d = B =\frac{ 5}{ 8} I_0$,
$C = \frac{ 3 }{ 8 } I_0$, and $E = C/2 $, where $I_0 = 7 \zeta(3) / 8 \pi^2 (T^*_c)^2 $. We see that $E >0$ and $B+E >C$.
 An elementary analysis then shows that the transition from $d$ to $s$ involves an intermediate phase in which the two orders mix with relative phase $\pm \pi/2$. This is long-thought $s \pm id$ state~\cite{hanke}.  The system chooses either $s+id$ or $s-id$ state and by this breaks time-reversal symmetry.
 The boundaries of this intermediate phase are set by
 $T_{s+id} = T^*_c (1 - \beta |\kappa - \kappa^* |)$, where
  $\beta = 6 \sqrt{3}/(u N_F)$.
  We emphasize again that the transition from $s$ to $d$ and the existence of the intermediate
   phase are both the consequences of the competition between intra-pocket and inter-pocket pairing
 (we recall that  $2 \Delta_s = c^{\dag}_{\uparrow} f^{\dag}_{\downarrow} + f^{\dag}_{\uparrow} s^{\dag}_{\downarrow}$,
$2 \Delta_d = c^{\dag}_{\uparrow} c^{\dag}_{\downarrow} - f^{\dag}_{\uparrow} f^{\dag}_{\downarrow}$).

 {\it The excitations:} Experiments on AFe$_2$Se$_2$ show~\cite{exp:AFESE,latest} that fermion excitations are fully gapped in the superconducting state.  In our theory
 this holds
  in the $s$-wave state
  and in
  the intermediate
    $s \pm id$ state.
    In the $d$-wave state, the excitation spectrum is given~\cite{suppl}:
    \begin{equation}
    \omega^2_{\pm} = |\Delta_d|^2 \cos^2 {2\theta} + \left(\sqrt{\xi^2 +  |\Delta_d|^2 \sin^2 {2\theta}} \pm \frac{\lambda}{\sin 2\theta}\right)^2 \, ,
    \label{n_1}
    \end{equation}
  where,
    we remind,
    $\xi = (\epsilon_c + \epsilon_f)/2 -\epsilon_F \approx v_F (k-k_F)$.    The dispersion $\omega_-$ has nodes (solutions
   with
   $\omega_{\pm}=0$)
    along the diagonal directions where $\cos 2 \theta =0$, as it should be for a $d$-wave superconductor. However, the nodal points are located in between $a$ and $b$ FSs, as shown in Fig.~\ref{fig:3}a, i.e., along the $a$ and $b$ FSs excitations do not have nodes. This is another consequence
      of the presence of inter-pocket pairing. We plot the dispersions  in $s$- and $d$-wave states in Fig.~\ref{fig:3}b.
      We furthermore see from \eqref{n_1}
       that nodes in the $d$-wave state exist only if $|\Delta_d| < \lambda$, otherwise the second term in the  r.h.s of  \eqref{n_1} does not vanish even when $\xi =0$.  The condition $|\Delta_d| = \lambda$ then sets the boundary of the {\it nodeless} $d$-wave state (we dubbed this state as $d'$ on Fig.~\ref{fig:1}). For a tetragonal system, realistic $\lambda$ are larger than $\Delta_d$, but we will see below that the presence of $d'$ state on the phase diagram is relevant to the structure of the nodes in the $d$-wave phase of  AFe$_2$Se$_2$.

\begin{figure}[h]
\begin{center}
\includegraphics[width=1.0\columnwidth]{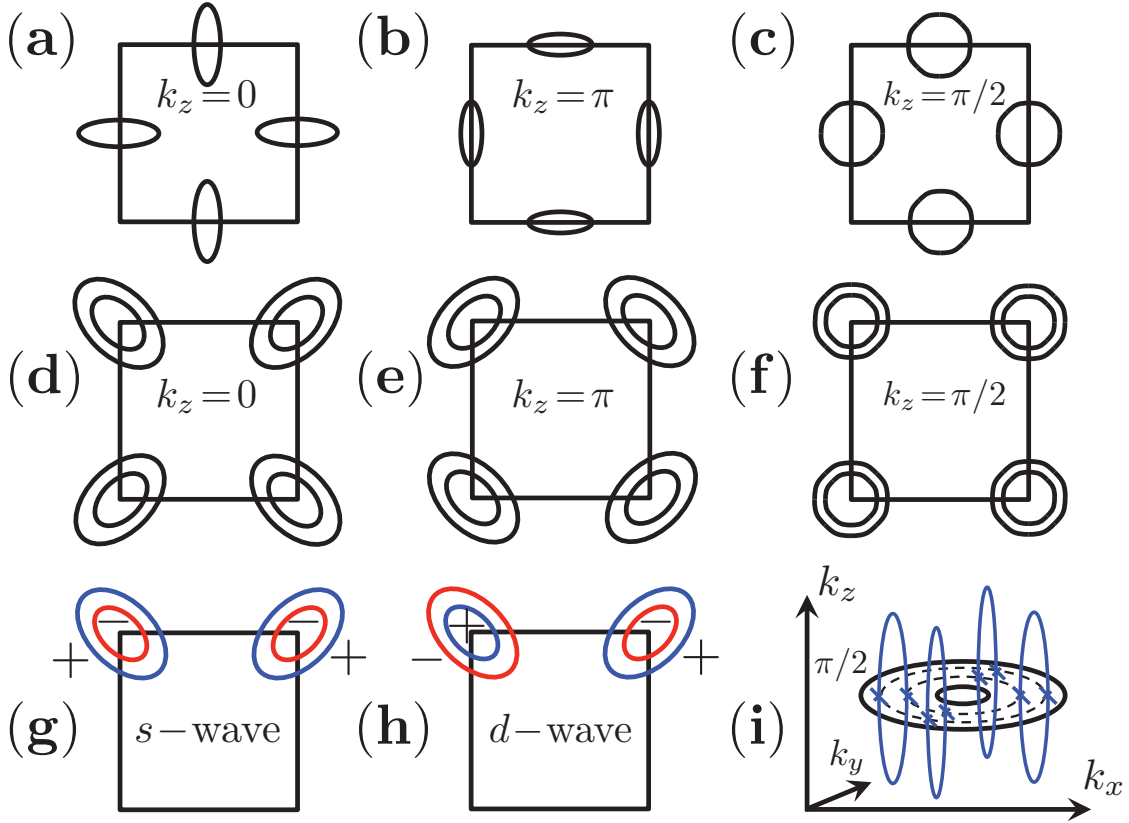}
\caption{The structure of electronic states and the superconducting gap in
  AFe$_2$Se$_2$ which have  body-centered tetragonal structure. Panels a-c -- electron pockets in the unfolded Brillouin zone for different $k_z$.
  Panels d-f -- same in the folded zone. Note that the two ellipses at each corner remain co-axial and rotate by $90^o$ between $k_z=0$ and $k_z =\pi$.
   The hybridized FSs at $k_z =\pi/2$ would be circular if one restricted with quadratic dispersion, as in \eqref{12} but acquire 4-fold variation due to higher-order
    terms in the dispersion. Panels g-h -- $s$-wave and $d$-wave gap structure near $k_z=0$ and $k_z=\pi$. Panel i -- the location of the nodes at $k_z \approx \pi/2$. The nodal points form vertical loops (only two are shown for clarity). If hybridized FSs at $\pi/2$ are two circles, the crosses extend and form lines in $(k_x,k_y)$ plane (dashed lines in the Figure).}
\label{fig:4}
\end{center}
\end{figure}

{\it Specifics of AFe$_2$Se$_2$:}
  The hybridization of electron pockets in AFe$_2$Se$_2$ is involved because
  of the body-centered tetragonal structure of these materials~\cite{exp:AFESE}.
 The two hybridized electron FSs differ by  $k_z = \pi$ and
are rotated by $\pi/2$ (see Fig.~\ref{fig:4}a-c
and Refs.\cite{mazin,inosov}).
 For $k_z =0$ and $k_z=\pi$,
 the FS in the folded zone consists of co-aligned ellipses
  (Fig.~\ref{fig:4}d-f),
  the pair near $(\pi,\pi)$
  at $k_z=0$ is identical to the one near
  $(-\pi,\pi)$
  at $k_z=\pi$.
   At $k_z = \pm \pi/2$, the pairs at
 $(\pi,\pi)$ and $(-\pi,\pi)$
   are identical already at the same $k_z$.
   $s$-wave and $d$-wave gaps differ in whether the gap on the larger
   ellipsis
   retains sign or changes sign between
   $k_z =0$ $k_z=\pi$ (Fig.~\ref{fig:4}g-h).
   Which of the two states
  is realized then depends on the behavior around $k_z = \pi/2$ where our 2D analysis is applicable~\cite{comm_2}.
   Using our 2D results, we find that a $d$-wave state, with intra-pocket pairing in terms of original fermions, wins at small $\kappa$, and
   an $s$-wave state, with intra-pocket pairing in terms of new, hybridized fermions, wins at large $\kappa$, and there is $s+id$ phase in between.
 In a $d$-wave state nodal points  exist near $k_z =\pi/2$ (again away from the FSs), but not near $k_z=0$ and $k_z=\pi$, where
  the FSs at the same $k_x, k_y$ in the folded zone are co-axial ellipses of different sizes, and hybridization
  can only cause minor variations of originally angle-independent gap
  (this behavior is the same as in $d'$ region in Fig.~\ref{fig:1}).
   Consequently
 the nodes in the $d$-wave state form vertical loops centered at $k_z = \pi/2$ (Fig.~\ref{fig:4}i).
  Vertical
   loop nodes, although of different kind,
    have been earlier suggested on phenomenological grounds~\cite{matsuda,dever}, but have not been computed microscopically.

 {\it {\bf Outlook}}--
In this work we argued that the pairing in recently discovered Fe-based superconductors AFe$_2$Se$_2$ (A = K, Rb, Cs) with only electron pockets must necessary include
  inter--band pairing correlations between fermions belonging to different pockets. We demonstrated that the interplay between intra-pocket and inter-pocket
   pairing leads, already within a "standard model", to a transition from $d$-wave pairing at small degree of hybridization to an $s$-wave pairing at larger hybridization. In terms of hybridized fermions $d$-wave is inter-pocket pairing and $s$-wave is intra-pocket pairing,  in terms of original fermions it is other way around. We found that
    the transformation from $d$ to $s$ is continuous and there is an intermediate, long-thought  $s\pm id$  mixed phase with broken  time reversal symmetry.
 Fermion excitations in $s$-wave and $s+id$ states are fully gapped,
  and $s$wave gap is of $s^{+-}$ type, with sign change between hybridized pockets. Such an $s$-wave
 gap should give rise to a spin resonance below $T_c$~\cite{mazin,ilya,tom}.
 The absence of nodes and the existence of a spin resonance are consistent with the data~\cite{exp:AFESE,resonance}. A $d$-wave state contains vertical loop nodes, the size of the loop
  depends on the ratio of hybridization and the gap amplitude, and is small for small hybridization.
 The normal state, $s$-wave, $d$-wave, and $s\pm id$ state all merge at the tetra--critical point on the phase diagram.
  We encourage photoemission studies of fermion excitations in  AFe$_2$Se$_2$ around $k_z =\pi/2$ and {\it away} from the FS to determine
    whether the pairing state is $s$-wave or $d$-wave, and studies of potential
 time reversal-symmetry breaking below $T_c$. We also encourage the analysis of the effects of  inter--band pairing in other Fe-based superconductors.

\section*{Acknowledgements}

We
are thankful to I. Mazin for numerous discussions, careful reading of the manuscript, and useful suggestions.
 We acknowledge helpful discussions with A. Bernevig, R. Fernandes,
 P. Hirschfeld, S. Graser, I. Eremin, K. Kuroki, D-H. Lee, S. Maiti,
  D. Scalapino, R. Thomale, and M. Vavilov. This work was supported by
University of Iowa (M.K.) and NSF-DMR-0906953 (A.C). A support from Humboldt foundation (A.C) is
gratefully acknowledged.
 M.K and A.C jointly identified the problem, performed the analysis and wrote the paper.
 Correspondence should be addressed to A.C.

\section{Supplementary Material}

In this supplement we present the calculations that were quoted in the main text.

\subsection{ Circular Fermi pockets}

In this case $\cos 2 \theta =0$, $\sin 2 \theta =1$ and the interaction Eqs.~\eqref{Hs_eff}, \eqref{Hd_eff} simplifies,
\begin{align}\label{H-O2}
H = -2 u_{s} [ J_-^{\dag} ]_{\sigma \sigma'} [ J_- ]_{\sigma' \sigma} -2 u_{d} [ \tilde{J}_+^{\dag} ]_{\sigma \sigma'} [ \tilde{J}_+ ]_{\sigma' \sigma}
\, .
\end{align}
With $u_s = u_d=u$ the inter-- and intra--band order parameters are degenerate at $\lambda=0$ as under the above $O(2)$ operations these two order parameters transform through each other.
At finite $\lambda$ the intra--band order should form first upon cooling as Cooper logarithms in this channel in contrast to inter-band pairing are unaffected by hybridization.
In terms of anomalous averages
$2 \Delta_s = \langle a_{\uparrow}^{\dag} a_{\downarrow}^{\dag} \rangle - \langle b_{\uparrow}^{\dag} b_{\downarrow}^{\dag} \rangle$,
$2 \Delta_d = \langle a_{\uparrow}^{\dag} b_{\downarrow}^{\dag} \rangle + \langle b_{\uparrow}^{\dag} a_{\downarrow}^{\dag} \rangle $
non-linear mean field equations obtained from \eqref{H-O2} reads
\begin{align}\label{D-s-MF}
1 = 2 u T \sum_n \frac{ \pi }{ \sqrt{ 4 u^2 \Delta_s^2 + \epsilon_n^2 } } \, .
\end{align}
The linearized mean field equation on $\Delta_d$ at finite $\Delta_s$ takes the form
\begin{align}\label{D-d-MF}
1 = 2 u T \sum_n \frac{ \pi }{ \sqrt{ 4 u^2 \Delta_s^2 + \epsilon_n^2 } }
\frac{  4 u^2 \Delta_s^2 + \epsilon_n^2 }{ 4 u^2 \Delta_s^2 + \epsilon_n^2 + \lambda^2 } \, .
\end{align}
It follows from comparison of \eqref{D-s-MF} and \eqref{D-d-MF} that once the $s$-wave order sets in,
the $d$-wave order is preempted.
This is a consequence of circular shape of Fermi pockets.

In the case of elliptical pockets and zero hybridization, the order parameter has opposite sign at two pockets, what implies a $d$-wave symmetry.
The hybridization, which  favors $s$-wave phase then competes with eccentricity which favors $d$-wave phase.

\subsection{Evaluation of the Ginzburg-Landau Functional}
To map a phase diagram in
 the $(\kappa,T)$-plane
we derive  the Ginzburg-Landau Functional (GLF).
We  follow  standard steps --  Hubbard-Stratonovitch decoupling in Cooper channel and integration over fermion fields,
and  obtain the expression for the free energy
\begin{align}\label{free}
F[ \Delta_s, \Delta_d] = -T\sum_{kn} \log \det  \!\left[ G^{-1}_{k,n} \right] + \frac{ |\Delta_s |^2 }{ u_s } + \frac{ |\Delta_d |^2 }{  u_d }  \, ,
\end{align}
where the inverse Green function is
\begin{align}
\label{matrix}
\!G^{-1}_{k,n}\!\! =\!\!\begin{bmatrix}
- (G^{a}_{kn,+})^{-1}
 & c \Delta_d \!+\! s \Delta_s  & 0 & -s \Delta_d \!+\! c \Delta_s \\
c \Delta_d^* \!+\! s \Delta_s^*   & (G^{a}_{kn,-})^{-1}& -s \Delta_d^*  \!+\! c \Delta_s^* & 0 \\
0 & -s \Delta_d  \!+\!  c \Delta_s & - (G^{b}_{kn,+})^{-1}  & - c \Delta_d  \!-\!  s \Delta_s \\
-s \Delta_d^*  \!+\! c \Delta_s^* & 0 & - c \Delta^*_d  \!-\!  s \Delta^*_s  & (G^{b}_{kn,-})^{-1}
\end{bmatrix}.
\end{align}
In the last equation the free Green functions are
$G^{a(b)}_{kn,\pm} = (\pm i \epsilon_n - \xi^{a(b)}_{ \pm k})^{-1} $,
$\xi_k^{a,b} =  E_k^{a,b} - \epsilon_F$.

Expansion of Eq.~\eqref{free} to fourth order yields GLF in the form
\begin{align}\label{GL1_1}
F_{GL}
= & A_s |\Delta_s|^2 + A_d |\Delta_d|^2
+  \frac{B_s}{2}    |\Delta_s|^4 +  \frac{B_d}{2}   |\Delta_d|^4
\notag \\
 & + C |\Delta_s|^2 |\Delta_d|^2
+  \frac{E_1}{2}( \Delta_s \Delta^*_d )^2 + \frac{E_2}{2} (\Delta^*_s \Delta_d )^2\, .
\end{align}

The expressions for $A_s$ and $A_d$ are
\begin{align}\label{AS}
A_{s} = \frac{ 1 }{  u } -  T \sum _{\vec{k}n}
\left[ s^2 ( G^a_- G^a_+ \! +\! G^b_- G^b_+ )
+
c^2 (G^a_- G^b_+   +  G^a_+ G^b_- )
\right]
\end{align}
and  $A_d$ is obtained from \eqref{AS} by exchanging $c \leftrightarrow s$.
Observe that Eq. \eqref{AS} contains both intra-- and intra--band contributions. They  correspond to
first and second terms in the square brackets, respectively.

The expressions for the other coefficients in terms of fermion propagators are
\begin{align}\label{Bs0}
B_s & =  T \sum _{\vec{k}n}
\Big[
s^4
\left( (G^a_-)^2  ( G^a_+)^2  +  (G^b_+)^2  ( G^b_-)^2 \right)
+
c^4
\left( (G^a_-)^2  ( G^b_+)^2  +  (G^a_+)^2  ( G^b_-)^2 \right)
 \\
 + &
2 c^2 s^2
\left( G^a_+ (G_-^a)^2 G^b_+  + G_+^a G^b_+ (G^b_-)^2  + G_-^a (G_+^a)^2 G_-^b
+ G^a_- (G^b_+)^2 G^b_- - 2 G_-^a G^a_+ G^b_- G^b_+
\right)
\Big] \, . \notag
\end{align}
The expression for the parameter $B_d$ is obtained from Eq.~\eqref{Bs0} by interchanging
$s \leftrightarrow c$.
Further,
\begin{align}\label{C0}
 C =  & T \sum _{\vec{k}n}
\Big[
(c^4 + s^4) \left( G_a^+ (G_a^-)^2 G_b^+  + G^a_+ G^b_+ (G_b^-)^2
+
(G_a^+)^2 G_a^-  G_b^-     + G_a^- (G_b^+)^2  G_b^- \right)
\notag \\
& +
2 c^2 s^2
\big(  (G_a^+)^2 (G_a^-)^2 + (G_a^-)^2 (G_b^+)^2
-
G_a^+ (G_a^-)^2  G_b^+   + (G_a^+)^2 (G_b^-)^2
\notag \\
& +
(G_b^+)^2 (G_b^-)^2   -  G_a^+ G_b^+ (G_b^-)^2
-
(G_a^+)^2  G_a^-  G_b^-   - G_a^- (G_b^+)^2 G_b^-
+ 4 G_a^+  G_a^- G_b^+ G_b^-
\big)
\Big]\, ,
\end{align}
\begin{align}\label{E0}
E_1 = E_2 = E
 = & T \sum _{\vec{k}n}
\Big[
-2 (c^4 + s^4) G_a^+  G_a^- G_b^+ G_b^-
\notag \\
& +
 c^2 s^2
\big(
(G_a^+)^2 (G_a^-)^2 + (G_a^-)^2 (G_b^+)^2
-2 G_a^+ (G_a^-)^2 G_b^+   + (G_a^+)^2 (G_b^-)^2
\notag \\
 & +
(G_b^+)^2  (G_b^-)^2   -2 G_a^+ G_b^+ (G_b^-)^2
-2
(G_a^+)^2 G_a^- G_b^-  -2  G_a^- (G_b^+)^2  G_b^-
\big)
\Big]\, .
\end{align}

The evaluation of GLF parameters simplifies in two limiting cases.
 At $T^*_c \ll \lambda$
 only the terms with propagators of particles
  from
  the same band,
   like
   the first two terms in Eq.~\eqref{Bs0}, are important in \eqref{AS}, \eqref{Bs0}, \eqref{C0} and \eqref{E0}.
  The parameters of the quadratic part of the GLF read
 \begin{align}\label{AS3}
A_{s(d)} = \frac{ 1 }{ u } -  f_{s(d)}(\kappa) \log \frac {\Lambda }{ T } \, ,
\end{align}
where
$f_s(\kappa) = \langle \sin^2 2 \theta_k \rangle$, $f_d = 1 - f_s$, $\sin^2 2 \theta_k = \kappa^2 / (\kappa^2 + \cos^2 2 \phi )$, $\cos \phi = k_x / k$ and $\langle \cdots \rangle = (2 \pi)^{-1} \oint d \phi \cdots$.
The critical $\kappa^* = 1 / \sqrt{3}$ is obtained from the condition $f_s(\kappa_*) = f_d(\kappa_*)$.

\subsection{The case $T^*_c \gg \lambda$}
For completeness, we consider the phase diagram for the case when
$T^*_c  \gg \lambda$ (we recall that $T^*_c$  is the transition temperature at the point where the lines $A_s =0$ and $A_d =0$ cross).
 We perform a perturbative evaluation of the GLF parameters in $\lambda/T^*_c \ll 1$.

Consider first the transition from normal to superconducting state.
 It follows from Eq.~\eqref{AS} that,  to the leading order in $\lambda/T^*_c$,
 $T_{c,d} = T_{c,s}$ independent on $\kappa$, like in $O(2)$ symmetric case  of circular Fermi pockets.
The finite difference $\delta A = A_s - A_d$ is obtained at the second order in $\lambda/T^*_c$,
\begin{align}\label{ASD1}
\delta A
\approx
\lambda^2 \kappa^{-2}
\langle \cos^2 2 \phi - \kappa^2 \rangle I_0 \, ,
\end{align}
where $I_0 = 7 \zeta(3) / 8 \pi^2 T^2 $.
As $\langle \cos^2 2 \phi \rangle = 1/2$, $\kappa_* = 1/ \sqrt{2}$.

To evaluate the other parameters in GLF
 perturbatively in $\lambda/T_c$ we write $\xi_{a,b} = \xi \pm \delta E$,
$\delta E \approx \lambda \kappa^{-1} \sqrt{ \kappa^2 + \cos^2 2 \phi } \ll \xi \sim T_c$ and expand
Green functions in $\delta E$.
Substituting the expansion in Eqs.~\eqref{Bs0}, \eqref{C0} and \eqref{E0} we obtain
$B_s^{(0)} = B_d^{(0)} = 2 I_0$,
$C^{(0)} = 4 I_0$, $E_1^{(0)} =  E_2^{(0)} = - 2 I_0$.
Because   $E_{1,2}^{(0)}<0$ we write the quartic part of GLF in the form
\begin{align}
F^{(4)} \approx  \frac{B^{(0)}}{2}( |\Delta_s|^2 +   |\Delta_d|^2)^2
+ ( C^{(0)}\! -\! B^{(0)}\! +\!  E^{(0)} )   |\Delta_s|^2 |\Delta_d|^2
+  \frac{E^{(0)}}{2} ( \Delta_s \Delta^*_d - \Delta^*_s \Delta_d )^2  \, .
\end{align}
\begin{figure}[h]
\begin{center}
\includegraphics[width=0.9\columnwidth]{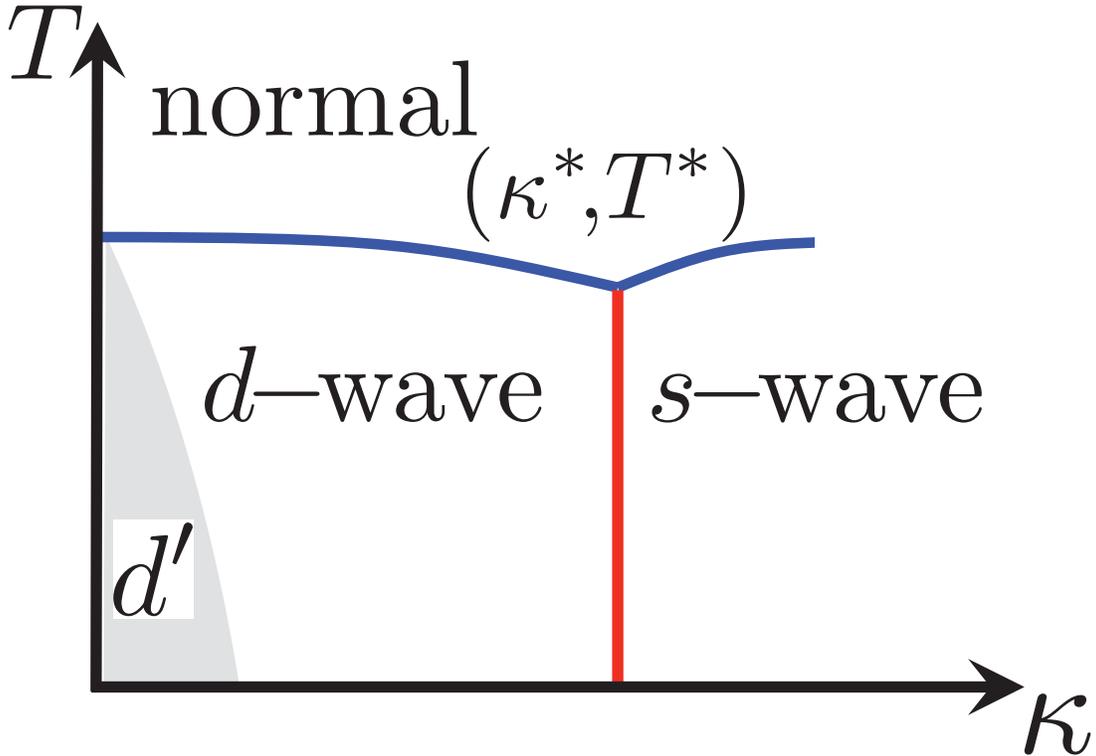}
\caption{Phase diagram in the $(\kappa,T)$--plane for $T^*_c \gg \lambda$.
 There is still a transformation from $d$-wave to $s$-wave gap symmetry as $\kappa$ increases, but the two phases are now separated by first-order transition.
 There is no intermediate $s +i d$ phase.  Within  the $d$-wave phase, there is still a region (dubbed as $d^{'}$) where the gap has no nodes.}
\label{fig1_1}
\end{center}
\end{figure}
The GLF is at minimum when $\Delta_s $ and  $\Delta_d$ are in phase, i.e., time-reversal symmetry is not broken in the strong coupling limit.
Still the leading order expansion is not sufficient to determine whether  the
intermediate phase of an $s + d$ symmetry exists because
$C^{(0)} - B^{(0)} +  E^{(0)} = 0$.
It turns out that at the second order GLF is still degenerate, $C^{(2)} - \sqrt{B^{(2)}_s B^{(2)}_d} + E^{(2)} = 0$, hence we
 higher order expansion. We obtained up to the fourth order in $\lambda/T_c$ found that $C  - \sqrt{B_s B_d} + E$ is positive:
\begin{align}
C  - \sqrt{B_s B_d} + E
& \approx
C^{(4)} - \frac{1}{2} ( B_s^{(4)} + B_d^{(4)} ) + E^{(4)}
= T \sum_{n}\int d \xi \frac{ 7 (\delta E)^4  }{  (\epsilon_n^2 + \xi^2)^4 }  \, .
\end{align}
This
 implies
  that at strong coupling,
    $s$-wave and $d$-wave phases are separated by first order transition line. We show the phase diagram for $T^*_c
 \gg
\lambda$  in Fig.~\ref{fig1_1}.


\begin{thebibliography}{5}

\bibitem{discovery}
 Y. Kamihara, T. Watanabe, M. Hirano, H.
Hosono, J. Am. Chem. Soc. \textbf{130}, 3296(2008).

\bibitem{review} A.F. Kemper, T.A. Maier, S. Graser, H-P. Cheng,
P.J. Hirschfeld and D.J. Scalapino, New J. Phys. \textbf{12},
073030(2010).

\bibitem{review_1}
P.J. Hirschfeld, M.M. Korshunov, and I.I. Mazin,
 Rep. Prog. Phys. \textbf{74}, 124508 (2011).

\bibitem{review_2} A.V. Chubukov, Annul. Rev. Cond. Mat. Phys. {\bf 3}, 13.1–13.36,  (2012).

\bibitem{review_3}  H.H. Wen and S. Li, Annu. Rev. Condens. Matter Phys.,  {\bf 2}, 121 (2011).

\bibitem{rahul} for the latest see R. Nandkishore, L. Levitov and A. Chubukov, Nature Phys. {\bf 8}, 158–163 (2012).

\bibitem{mazin_1}  I. I. Mazin, D. J. Singh, M. D. Johannes, and M.
H. Du, Phys. Rev. Lett. \textbf{101}, 057003 (2008).

\bibitem{kuroki_1}
K. Kuroki, S. Onari, R. Arita, H. Usui, Y. Tanaka, H. Kontani, and H. Aoki, Phys. Rev. Lett. \textbf{101}, 087004 (2008).

\bibitem{cvv}  A. V. Chubukov, M. G. Vavilov, A. B. Vorontsov,
Phys. Rev. B \textbf{80}, 140515(R)(2009).

\bibitem{exp:AFESE} J. Guo, S. Jin, G. Wang, S. Wang, K. Zhu, T. Zhou, M. He, and X. Chen, Phys. Rev. B {\bf 82}, 180520(R) (2010).

\bibitem{latest} For most recent results see Y. Liu, Z. C. Li, W. P. Liu, G. Friemel, D. S. Inosov, R. E. Dinnebier, Z. J. Li, C. T. Lin,
 arXiv:1201.0902 and referennces therein.

\bibitem{exp:AFESE_ARPES} T. Qian, X.-P. Wang, W.-C. Jin, P. Zhang, P. Richard, G. Xu, X. Dai, Z. Fang, J.-G. Guo, X.-L. Chen, H. Ding, Phys. Rev. Lett. \textbf{106}, 187001 (2011).




\bibitem{graser_11}  T.A. Maier, S. Graser, P. J. Hirschfeld, and D. J. Scalapino,
Phys. Rev. B \textbf{83}, 100515(R) (2011)

\bibitem{das} T. Das and A. V. Balatsky, Phys. Rev. B 84, 014521 (2011);
Phys. Rev. B \textbf{84}, 115117 (2011)

\bibitem{dhl_AFESE} F. Wang, F. Yang, M. Gao, Z.-Y. Lu, T. Xiang and D.-H. Lee, Europhys. Lett. \textbf{93} 57003
(2011).

\bibitem{maiti}
S. Maiti, M. M. Korshunov, T. A. Maier, P. J. Hirschfeld, and A. V. Chubukov, Phys. Rev. B {\bf 84}, 224505 (2011); Phys. Rev. Lett. {\bf 107}, 147002 (2011).

\bibitem{mazin} I.I. Mazin, Phys. Rev. B \textbf{84}, 024529
(2011).

\bibitem{si_11}  R. Yu, P. Goswami, Q. Si, P. Nikolic, J.-X. Zhu, arXiv:1103.3259.

\bibitem{bernevig_11}
C. Fang, Y.-L. Wu, R. Thomale, B. A. Bernevig, J. Hu, Physical
Review X \textbf{1}, 011009 (2011)

\bibitem{kontani_last} T. Saito, S. Onari, and H. Kontani \prb \textbf{83}, 140512(R)
(2011).




\bibitem{Zhang} Y. Zhang, L. X. Yang, M. Xu, Z. R. Ye, F. Chen, C. He, J. Jiang, B. P. Xie, J. J. Ying, X. F. Wang, X. H. Chen, J. P. Hu, D. L. Feng, Nature Materials {\bf 10}, 273-277 (2011).
\bibitem{Zhou} D. Mou, S. Liu, X. Jia, J. He, Y. Peng, L. Zhao, Li Yu, G. Liu, S. He, X. Dong, J. Zhang, H. Wang, C. Dong, M. Fang, X. Wang, Q. Peng, Z. Wang, S. Zhang, F. Yang, Z. Xu, C. Chen, and X. J. Zhou, Phys. Rev. Lett. {\bf 106}, 107001 (2011).


\bibitem{Wang} X.-P. Wang, T. Qian, P. Richard, P. Zhang, J. Dong, H.-D. Wang, C.-H. Dong, M.-H. Fang, H. Ding
 Europhysics Letters {\bf 93}, 57001 (2011).



\bibitem{resonance} J. T. Park, G. Friemel, Y. Li, J.-H. Kim, V. Tsurkan, J. Deisenhofer, H.-A. Krug von Nidda, A. Loidl, A. Ivanov, B. Keimer, and D. S. Inosov, Phys. Rev. Lett. 107, 177005 (2011).

\bibitem{efetov} M. Einenkel and K. B. Efetov, Phys. Rev. B {\bf 84}, 214508 (2011).

\bibitem{Zhang_1}
X. Dai, Z. Fang, Y. Zhou , F-C Zhang,
 Phys. Rev. Lett. 101, 057008 (2008).

 \bibitem{mazin_3} I.I. Mazin, M.D. Johannes, L. Boeri, K. Koepernik, D.J. Singh, Phys. Rev. B {\bf 78}, 085104 (2008).


\bibitem{inosov} J. T. Park, D. S. Inosov, A. Yaresko, S. Graser, D. L. Sun, Ph. Bourges, Y. Sidis, Yuan Li, J.-H. Kim, D. Haug, A. Ivanov, K. Hradil, A. Schneidewind, P. Link, E. Faulhaber, I. Glavatskyy, C. T. Lin, B. Keimer, V. Hinkov,   Phys. Rev. B {\bf 82}, 134503 (2010)

\bibitem{comm}  For circular pockets, the interaction Hamiltonian  \eqref{14} is expressed via the square of the total charge density, $n = c^\dagger_\alpha c_\alpha + d^\dagger_\alpha d_\alpha$, total ${\bf S}^2$, where ${\bf S} = (1/2) (c^\dagger_\alpha c_\beta + d^\dagger_\alpha d_\beta) {\bf \sigma}_{\alpha \beta}$, and $O(2)$ and $SU(2)$-invariant
     combination ${\tilde n}^2$, where ${\tilde n} = c^\dagger_\alpha d_\alpha - d^\dagger_\alpha c_\alpha$, as
     $H = (U) n^2/2 + J^{'} {\tilde n}^2/2 + 2J {\bf S}^2$.  The interactions $u_i$ in \eqref{14}  are $u_1 = U-3J, u_2 = -2J-J^{'}, u_3 = J^{'}, u_4 = U-3J$.  Then $u_4-u_3 = u_1+u_2 = U-3J-J^{'}$.

\bibitem{suppl} See Supplementary material.

\bibitem{hanke} Our scenario for $s+ id$ state is diiferent from the one proposed by  C. Platt,  R. Thomale, C. Honerkamp, S-C. Zhang, and W. Hanke,  arXiv:1106.5964. They found a sliver of $s+i d$ state in systems with hole and electron pockets, due to competition between intra-pocket $s$ and $d$ pairing.

\bibitem{comm_2}  The pairing problem at $k_z = \pi/2$ generally involves four sets of fermions
at $(\pi,0, \pm \pi/2$ and $(0,\pi, \pm \pi/2)$, of which two pairs hybridize and all four fermions interact with each other.
 This problem is equivalent to our 2D problem if the interaction is peaked at $k_z = \pi$ because then
  the pairing problem  decouples into two independent sets, each contains a  pair of fermions, like in our 2D analysis.

\bibitem{matsuda}
M. Yamashita, Y. Senshu, T. Shibauchi, S. Kasahara, K. Hashimoto, D. Watanabe, H. Ikeda, T. Terashima, I. Vekhter, A. B. Vorontsov, Y. Matsuda, Phys. Rev. B {\bf 84}, 060507(R) (2011).

\bibitem{dever} I. I. Mazin, T. P. Devereaux, J. G. Analytis, Jiun-Haw Chu, I. R. Fisher, B. Muschler, and R. Hackl,  Phys. Rev. B 82, 180502(R) (2010).

\bibitem{ilya} M.M. Korshunov, and I. Eremin, Phys. Rev. B {\bf 78}, 140509(R) (2008).

\bibitem{tom} T.A. Maier, and D.J. Scalapino, Phys. Rev. B {\bf 78}, 020514(R)
(2008).


\end{thebibliography}
\end{document}